\documentclass[runningheads]{llncs}

\usepackage[T1]{fontenc}
\usepackage{graphicx}
\usepackage{mathtools}
\usepackage{amssymb}
\usepackage{multirow}
\usepackage{mathtools}
\usepackage{pifont}

\usepackage{booktabs}
\usepackage{makecell}
\usepackage{threeparttable}[flushleft]
\usepackage{xcolor}

\usepackage{hyperref}
\usepackage{cleveref} % after hyperref

\usepackage{url}
\usepackage{color}

\hypersetup{
    colorlinks,
    linkcolor={blue},
    citecolor={blue},
    urlcolor={blue}
}

\usepackage{algorithm}
\usepackage{algorithmic}

\definecolor{mygreen}{rgb}{0.0, 0.5, 0.0}
\definecolor{myred}{rgb}{0.55, 0.0, 0.0}

\newcommand{\cmark}{\color{mygreen}{\ding{51}}}%
\newcommand{\xmark}{\color{myred}{\ding{55}}}%
\newcommand{\ourmod}{X-SiT}

\begin{document}

\title{\ourmod: Inherently Interpretable Surface Vision Transformers for Dementia Diagnosis}
\titlerunning{\ourmod}

\author{Fabian Bongratz\inst{1,2*},
Tom Nuno Wolf\inst{1,2*},
Jaume Gual Ramon\inst{1}, \\
Christian Wachinger\inst{1,2}
}

\authorrunning{F.~Bongratz, T.N.~Wolf, et al.}
\institute{Lab for AI in Medical Imaging, Technical University of Munich, Munich, Germany
\\
\email{\{fabi.bongratz,tn.wolf\}@tum.de}
\and
Munich Center for Machine Learning, Munich, Germany
}

\renewcommand{\thefootnote}{}% Remove number
\footnotetext{*Equal contribution.}%
\addtocounter{footnote}{-1}% Prevent counter from increasing
\renewcommand{\thefootnote}{\arabic{footnote}}% Restore numbering

\maketitle              % typeset the header of the contribution
\begin{abstract}
Interpretable models are crucial for supporting clinical deci-sion-making, driving advances in their development and application for medical images. However, the nature of 3D volumetric data makes it inherently challenging to visualize and interpret intricate and complex structures like the cerebral cortex. Cortical surface renderings, on the other hand, provide a more accessible and understandable 3D representation of brain anatomy, facilitating visualization and interactive exploration. Motivated by this advantage and the widespread use of surface data for studying neurological disorders, we present the eXplainable Surface Vision Transformer~(\ourmod). This is the first inherently interpretable neural network that offers human-understandable predictions based on interpretable cortical features. 
As part of \ourmod, we introduce a prototypical surface patch decoder for classifying surface patch embeddings, incorporating case-based reasoning with spatially corresponding cortical prototypes. 
The results demonstrate state-of-the-art performance in detecting Alzheimer's disease and frontotemporal dementia while additionally providing informative prototypes that align with known disease patterns and reveal classification errors. 

\end{abstract}

\section{Introduction}

Explainability, i.e., making an automated decision-making process understandable for humans, is a fundamental prerequisite for applications in clinical environments~\cite{Yang2022unbox}. Yet, deep neural networks do not fulfill this property without further modification due to their complex information-processing architecture.
Therefore, previous endeavors have focused on developing explainable methods for clinical reports, tabular measurements, video data, and medical images~\cite{gallee2023interpretable,mohammadjafari2021using,PahuddeMortanges2024orchestrating,wolf2023don}.

Cortical surface analysis, however, relies on non-Euclidean data for understanding various neurological and psychiatric conditions~\cite{deChastelaine2023cthCognitive,Chouliaras2023differentialDiagnosis,Pang2023geometricConstraints}. Specifically, triangular meshes are commonly used to represent the cortical sheet and morphological measurements thereof~\cite{Dale1999corticalI,Fischl1999corticalII}.
A central advantage of surface-based representations compared to 3D volumetric images is that
they preserve spatial relations and topological properties of the cortical sheet, crucial for faithful explanations. Moreover, surfaces are well-suited for 3D visualization; they support interactive exploration, where specific areas of interest can be examined from different angles and depths, providing a more immersive user experience compared to traditional 2D images. Plotting and interpreting surface-based biomarkers, e.g., cortical thickness, is also straightforward.
As these features are obtained after considerable processing, they also reduce potential sources of bias, e.g., inhomogeneities in image intensities~\cite{Nyul2000standardization}. 

\noindent
\textbf{Related work.}
Dedicated neural networks were previously developed for cortical surfaces, based on graph-convolutions~\cite{Gopinath2022gcn,Zhao2019sunet} and transformers~\cite{Cheng2022spherical,dahan22sit}. While the receptive field of graph-convolutional kernels is limited, transformer-based models can capture long-range dependencies across the cortical sheet.
Unfortunately, the decision-making of transformers can only be understood vaguely by the attention weights~\cite{dahan22sit}, which are considered unreliable explanations~\cite{kashefi2023explainability,komorowski2023towards}. 
Technically most related to our work, an extension of the Vision Transformer (ViT)~\cite{dosovitskiy2021vit} with a neural tree decoder~(ViT-NeT)~\cite{kim22vitnet} was proposed to explain the decision-making of transformers in computer vision. However, ViT-Net has inherent flaws when it comes to processing registered cortical surface data. There are no spatial restrictions on the matching of prototypes; while a bird can be anywhere in an image, brain regions always align to a template. Moreover, the sequential traversal of local binary decisions prevents a global comparison of structural brain patterns.
Most interpretable methods in neuroimaging are designed for images~\cite{Munroe2024interpretableNeuroReview}. Notable exceptions use saliency maps~\cite{Azcona2020interpretation,Spitzer2022interpretable}, which do not explain the actual reasoning process of the neural network~\cite{rudin2019stop}.

\begin{figure}[t]
    \centering
    \includegraphics[width=0.9\textwidth]{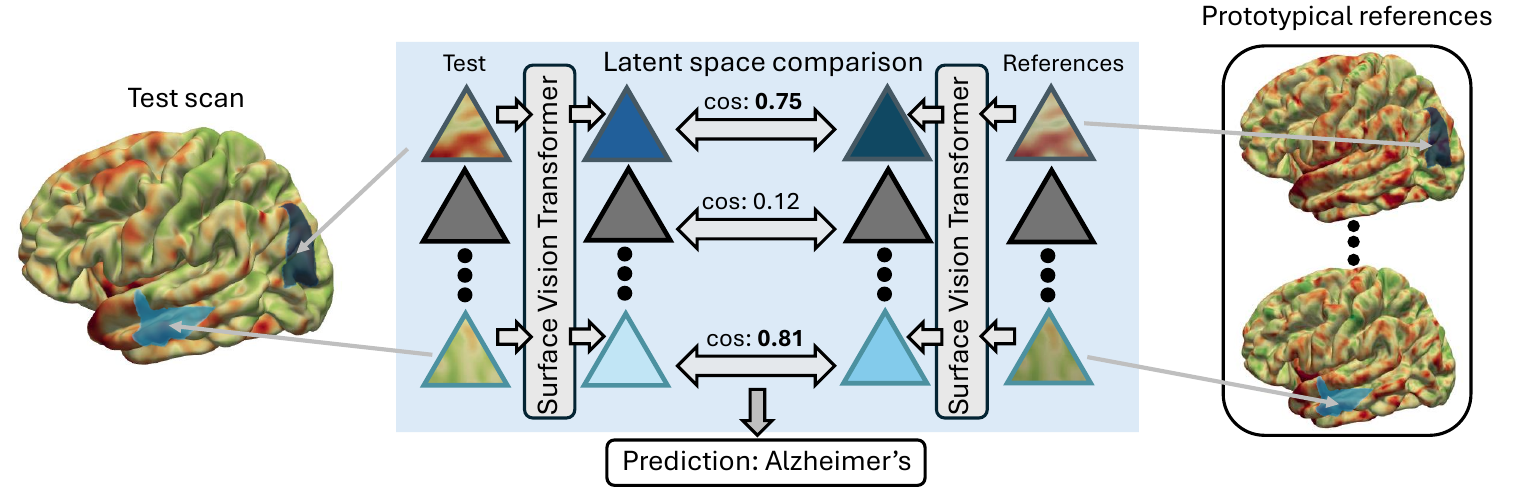}
    \caption{Case-based reasoning with \ourmod. X-SiT learns discriminative prototypes of cortical surfaces; patches are colored in blue and surfaces represent cortical thickness. 
    }
    \label{fig:cbm}
\end{figure}

\noindent
\textbf{Contribution.}
We present the eXplainable Surface Vision Transformer~(X-SiT), an \emph{inherently interpretable} transformer-based neural network for the classification of cortical surface data. 
X-SiT leverages the cosine similarity to learned \emph{corresponding cortical prototypes}, which represent a specific cohort, for the decision-making; see \Cref{fig:cbm} for a visualization. We evaluated X-SiT for detecting two common forms of dementia, Alzheimer's disease~(AD) and frontotemporal dementia~(FTD), demonstrating competitive performance to state-of-the-art models. We show that the prototypes learned by X-SiT closely match the discriminative disease patterns from the literature, and we demonstrate their benefit for comprehending the model's predictions both locally and globally.

\section{Methods}

\begin{figure}[t]
    \centering
    \includegraphics[width=0.9\textwidth]{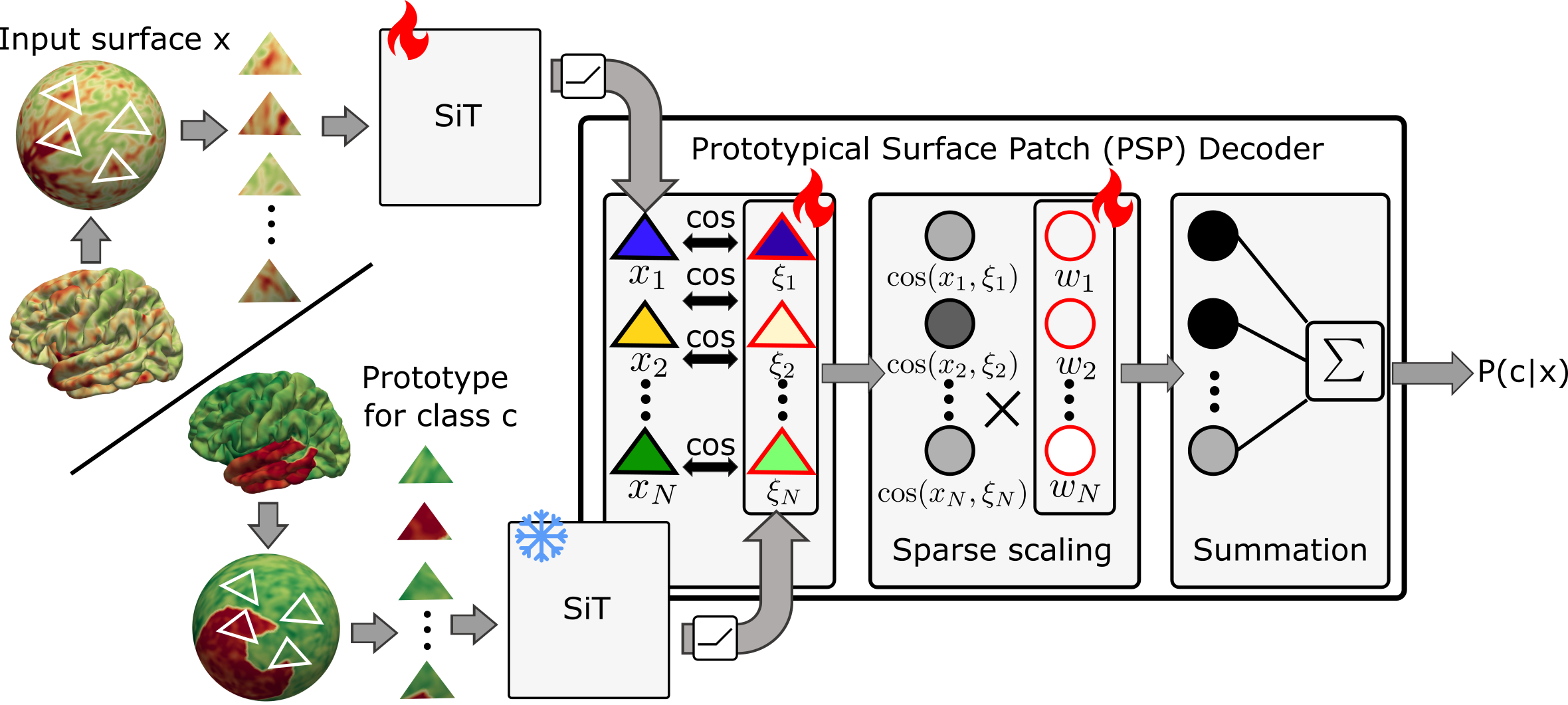}
    \caption{Architecture of the eXplainable Surface Vision Transformer~(X-SiT). 
    }
    \label{fig:figtech}
\end{figure}

We illustrate the architecture of our X-SiT model in \Cref{fig:figtech}. In the following, we describe its individual building blocks, namely the surface encoder and the prototypical surface patch (PSP) decoder. On a high level, the X-SiT takes a registered cortical surface $\mathbf{x}$ as input and computes a probability $P(c\mid \mathbf{x})$ of the surface $\mathbf{x}$ belonging to a certain target class $c$.

\subsection{Surface Transformer Encoder} \label{sec:sit-encoder}
We use the Surface Vision Transformer~(SiT)~\cite{dahan22sit} to encode the input surface, given as triangular mesh with spherical topology. 
Throughout this work, we assume these surfaces were pre-registered to a template, therefore consisting of a fixed number of $|V|$ vertices. 
Each vertex comprises a set of $F$ local features, such as cortical thickness, sulcal depth, and curvature. 
We partition the surface into a fixed set of $N$ triangular, non-overlapping surface patches. Hence, we obtain a sequence of $N$ patches, each comprising $M$ vertices. These patches serve as the input to the SiT encoder, yielding patch-wise latent embeddings.  
Specifically, we convert each input patch of dimension $M\times F$ into a latent vector of dimension $D$, corresponding to a functional mapping $f_{\text{SiT}} : \mathbb{R}^{N \times M \times F} \mapsto \mathbb{R}^{N \times D}$.
The resulting sequence of encoded patches, which we denote as $x = (x_1, \ldots, x_N)$, %$\{x_1, \ldots, x_N\}$, 
serves as input to X-SiT's decoder, which we will explain next.

\subsection{PSP Decoder} \label{sec:proto}
To decode the patches into a probability $P(c\mid \mathbf{x})$ of the input $\mathbf{x}$ belonging to a certain class $c$ --- in an interpretable manner, we propose a new prototypical surface patch~(PSP) decoder. 
The core idea is to compare the encoded patches to learned prototypes, which consist of brain regions from exemplary cases representing the target class $c$, as illustrated in \Cref{fig:cbm}. 
The underlying principle, case-based reasoning~\cite{chen2019looks}, allows us to reason as ``this part in the input looks like that part from the training set'', the most human-comprehensible reasoning process~\cite{jeyakumar2020can,kim2023help,nguyen2021effectiveness}. However, the nature of cortical surfaces requires \emph{corresponding prototypes}, i.e., matching brain regions, to allow for reasoning such as ``this cortical region in the input looks like that corresponding cortical region from the training set''.
This is in contrast to image-based prototypical networks~\cite{chen2019looks,kim22vitnet,nauta2023pipnet}, where the presence of prototypes is typically investigated across the entire image, but it
ensures the comparisons made between cortical inputs and examples from the training set are clinically and anatomically meaningful. 
By allowing the model to take individual patches from different training cases, X-SiT can learn prototypical disease patterns that a single case could hardly cover.

\noindent
\textbf{Prototype similarity.}
Like the input surface, we encode the prototypes with our SiT encoder, cf. \Cref{fig:figtech}. Thereby, we obtain a second, prototypical sequence of encoded patches, which we denote by $\xi = (\xi_1, \ldots, \xi_N)$. Based on the correspondences, two encoded patches $x_i$ and $\xi_i$ with the same index $i$ can be compared based on the cosine similarity of their latent representations: 
\begin{align}
    \cos(x_i, \xi_i) = \frac{\sum^D_{j=1} x_{i,j} \xi_{i,j}}{\sqrt{\sum^D_{j=1} x^2_{i,j}} \sqrt{\sum^D_{j=1} \xi^2_{i,j}}}.
\end{align}
Since the prototypical patches represent the target class locally, the cosine similarity provides local evidence for or against characteristic structural patterns being present in the input surface. To further improve interpretability, we rectify all input embeddings to be positive via a ReLU function, prohibiting an increase of class probability in the absence of features. Therefore, only positive evidence of the target class is considered, as opposed to ``no evidence'' when $\cos(x_i, \xi_i) \approx 0$.

\noindent
\textbf{Sparse scaling.}
While the cosine similarity indicates the presence of class-related structural patterns, it does not distinguish between important and less important regions. To equip X-SiT with the capability to re-weight the importance of individual regions, we scale the cosine similarities with weights $w=\{w_1, \ldots, w_N\}$, such that $\sum_{i=1}^N w_i=1$. We enforce sparsity by suppressing the contribution of below-average regions, i.e., weights are set to zero iff $w_i<1/N$.
The rationale behind the sparsification is to encourage the model to focus primarily on discriminative brain regions. 
Moreover, sparse decision-making is generally known to improve model interpretability~\cite{rudin2022interpretable,tibshirani1996regression}.

\noindent
\textbf{Summation and final prediction.}
Finally, we compute the class probability $P(c\mid \mathbf{x})$ as the weighted sum of cosine similarities: 
\begin{equation}\label{eq:proba}
    P(c\mid \mathbf{x}) = \sum_{i=1}^{N} w_i \cos(x_i, \xi_i).
\end{equation}
Given that the sum of the weights equals one~(cf.~definition above), a high probability $P(c\mid\mathbf{x})\approx 1$, indicates that the input surface resembles the prototypes in all relevant regions, i.e., regions where $w_i>0$.

\subsection{Training}
We train the SiT encoder, the sparse scaling weights, and the prototypes with binary cross-entropy end-to-end, as indicated by the flames in \Cref{fig:figtech}. Therefore, the SiT maps the input surface $\mathbf{x}$ to an optimal latent space for comparison with the prototypes.
Every five epochs, we replace each prototypical patch --- specifically, its latent representation $\xi_i$
--- with the most similar patch from all training instances at the prototype’s location.
Thus, the prototypes correspond to actual training samples without impairing the stability of the training process.

\section{Results and Discussion}

\subsection{Experimental Setting}\label{sec:setting}
We consider two clinical diagnostic tasks to evaluate the performance of the developed X-SiT model, namely Alzheimer's disease~(AD) diagnosis and detection of frontotemporal dementia~(FTD). We compare our approach to non-interpretable state-of-the-art models for registered meshes, i.e., SpiralNet$^{++}$~\cite{Gong2019spiral}, Spherical U-Net~(S-UNet)~\cite{Cheng2022spherical}, and the original SiT~\cite{dahan22sit}. Additionally, we implemented an inherently interpretable surface transformer, SiT-NeT, following the architecture and design of ViT-NeT~\cite{kim22vitnet}.
We tuned hyper-parameters\footnote{Search space included learning rate, weight decay, number of conv/transformer blocks, latent channels, and model-specific parameters such as sequence length for SpiralNet$^{++}$ and number of attention heads for SiT/SiT-NeT/X-SiT. 
}
for all models, based on the balanced accuracy (Bacc) on the validation set. We trained each model with five random initializations and AdamW~\cite{loshchilov2018decoupled}, using inversely proportional class-frequency-based loss weights to mitigate class imbalance; all reported values are the mean and standard deviation across these runs.

\begin{table}[t]
  \setlength{\tabcolsep}{3pt}
\small
\centering
\caption{Statistics of the datasets used for our experiments. We apply our model to diagnosing Alzheimer's disease~(AD) and frontotemporal dementia~(FTD). We report the size of each cohort, the percentage of female subjects~(\%F), and the age~(mean\textpm std.).}
\begin{tabular}{l@{\hspace{15pt}} lrrcc @{\hspace{15pt}}lrccc}
\toprule

& 
\multicolumn{5}{c}{Task 1: AD diagnosis} & 
\multicolumn{5}{c}{Task 2: FTD diagnosis}
\\
\cmidrule(l{0.3em}r{1.6em}){2-6}
\cmidrule(lr){7-11}

Split &

\makecell[l]{Dataset}  & 
CN &
AD &
\%F & 
Age

&

\makecell[l]{Dataset}  & 
CN &
FTD &
\%F & 
Age
\\

\midrule

Train &
\multirow{3}{*}{ADNI} &
334 & 192 & 49.2 & $74.2 \pm 6.6$ &
\multirow{3}{*}{\makecell[l]{ADNI/\\ NIFD}} & 
392 & 44 & 50.7 & $71.3 \pm 8.1$
\\

Val & 
&
56 & 32 & 51.1 & $73.8 \pm 6.7$ &
&
61 & 10 & 49.3 & $70.4 \pm 8.1$  \\

Test 
&
& 55 & 32 & 49.4 & $74.3 \pm 6.5$  &
&
64 & 13 & 50.6 & $69.6 \pm 8.1$ 
\\
\bottomrule

\end{tabular}
\label{tab:dataset_statistics}
\end{table}

\subsection{Data and Preprocessing} \label{sec:data}
We used public data from the Alzheimer's disease neuroimaging initiative~(ADNI)\footnote{\url{https://adni.loni.usc.edu}} and the frontotemporal lobar degeneration neuroimaging initiative~(NIFD)\footnote{\url{https://memory.ucsf.edu/research-trials/research/allftd}}. 
We used only baseline scans to avoid subject bias and data leakage. We split the data into training, validation, and test sets, stratified according to age, sex, and diagnosis; see \Cref{tab:dataset_statistics} for an overview. We focused on subjects diagnosed with AD from ADNI and FTD patients (behavioral variant; bvFTD) from NIFD, comparing them to cognitively normal~(CN) reference cohorts. 
Due to the low number of CN samples in the NIFD dataset~($n=72$), we combined the CN groups from both studies for the FTD diagnosis task. We processed all T1w MR images with FreeSurfer~(v7.2)~\cite{fischlFreeSurfer2012}, obtaining vertex-wise cortical thickness, curvature, and sulcal depth maps. Subsequently, we registered the data to the FsAverage template, represented as 6th-order icosphere (40,962 vertices). 
We concatenated the vertices of left and right hemispheres as inputs for all models.

\begin{table}[t]

\centering
\setlength{\tabcolsep}{1.8pt}
\renewcommand\bfdefault{b}% rather than bx
\begin{threeparttable}
    \centering
    \small
    \caption{Classification results based on validation (V) and test (T) sets for the two considered diagnostic tasks, i.e., Alzheimer's disease~(AD) and frontotemporal dementia~(FTD) vs.~cognitively normal, respectively. We report mean\textpm std. of five models with different weight initialization. We highlight \textbf{best} and \emph{second} test results, and we indicate the model's interpretability.}
    \begin{tabular}{lc lll lll }
    \toprule
    && \multicolumn{3}{c}{AD diagnosis} & \multicolumn{3}{c}{FTD diagnosis} \\
    \cmidrule(lr){3-5}
    \cmidrule(lr){6-8}
         Model & Interp. & Bacc~(V) & Bacc (T) & F1~(T) & Bacc~(V) & Bacc (T) & F1~(T)  \\
         \midrule
         SpiralNet$^{++}$~\cite{Gong2019spiral} & \xmark & 73.3\textpm1.3 & 70.3\textpm0.7 & 69.0\textpm0.8 & 88.5\textpm3.4 & 79.5\textpm0.8 & 72.3\textpm2.5 \\
         S-UNet~\cite{Zhao2019sunet}\tnote{1} & \xmark & 65.0\textpm2.6 & 57.6\textpm3.5 & 57.2\textpm3.5 & 76.9\textpm5.2 & 73.3\textpm5.3 & 73.0\textpm7.0\\
         SiT~\cite{dahan22sit} & \xmark & 85.3\textpm2.8 & \textbf{80.5\textpm0.5} & \emph{80.5\textpm0.9} & 85.3\textpm2.9 & \textbf{80.6\textpm1.8} & \textbf{74.3\textpm4.4} \\
         SiT-NeT\tnote{2} & \cmark & 75.0\textpm3.6 & 68.5\textpm3.6 & 68.6\textpm3.6 & 85.3\textpm4.6 & 72.6\textpm5.0 & 65.3\textpm6.8 \\
         \ourmod\ (ours) & \cmark & 84.5\textpm1.4 & \emph{80.2\textpm2.0} & \textbf{80.6\textpm3.1} & 87.5\textpm3.0 & \emph{79.6\textpm1.6} & \emph{74.0\textpm2.6} \\
         \bottomrule
    \end{tabular}
     \begin{tablenotes}
        \scriptsize
        \item[1] We used only the encoder of S-UNet, with adaptive average pooling and a classification head.
        \item[2] We adapted the ViT-Net model~\cite{kim22vitnet} for cortical surface data.
    \end{tablenotes}
    \label{tab:results}
    \end{threeparttable}
\end{table}

\subsection{Classification Performance}\label{sec:results-classification}
We compare the classification performance of \ourmod\ to the implemented baseline methods in \Cref{tab:results}. Notably, for AD detection , X-SiT and SiT outperform the graph convolution-based models~(S-UNet and SpiralNet$^{++}$) by a large margin~(+10\% Bacc on the test set). This is likely due to the ability of transformers to capture the widespread patterns of cortical degeneration in AD~\cite{Singh2006corticalthinningalzheimers}, whereas graph convolutions are inherently limited to aggregating structural information from local neighborhoods. For the diagnosis of FTD, this advantage seems less significant; SpiralNet$^{++}$ achieves the highest Bacc of 88.5\% on the validation set. However, X-SiT and SiT are again at the forefront on the test set (Bacc of 80.6 and 79.6, respectively), indicating better generalization and less sensitivity to hyper-parameters tuned on the validation set. Overall, the results presented in \Cref{tab:results} demonstrate that X-SiT maintains competitive performance compared to its direct non-interpretable baseline, SiT, while also providing valuable explanations~(see \Cref{sec:results-interpretablility} in the following). Notably, the proposed PSP decoder in X-SiT consistently outperformed the neural tree decoder in SiT-NeT.

\begin{figure}[t]
    \centering
    \includegraphics[width=0.8\textwidth]{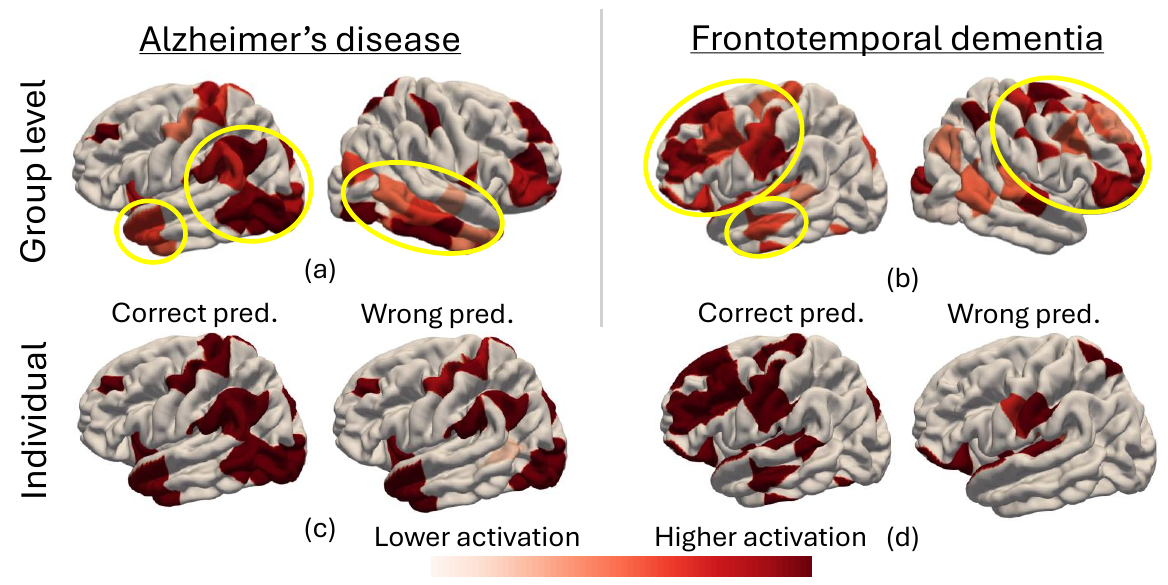}
    \caption{
    (a) and (b) show mean activation (weighted similarity to prototypes) in red across correct test set classifications of the dementia class. Higher activation implies a higher contribution to the detection of the respective disease. Yellow ellipses indicate the hallmark regions of the respective dementia type: (a) temporal lobe and temporoparietal junction for Alzheimer's disease and (b) frontal and temporal regions for frontotemporal dementia. (c) and (d) show model activation for individual predictions.
    }
    \label{fig:meanActivations}
\end{figure}

\subsection{Interpretability}\label{sec:results-interpretablility}
\textbf{Group level.}
\ourmod\ provides \emph{global} explanations on the group level that help understanding the model's general decision-making. In \Cref{fig:meanActivations}~(a) and (b), we show the weighted patch-wise similarity, $w_i\cos (x_i, \xi_i)$, of input patches to real-instance prototypes, averaged across the respective disease group in the test set. Specifically, \Cref{fig:meanActivations}~(a) indicates that \ourmod, trained for AD detection, focuses on regions that are actually characteristic of AD: the temporal lobe and the temporoparietal junction~\cite{Singh2006corticalthinningalzheimers}. Additionally, we found contributions from the frontal and occipital lobes, both of which were previously associated with AD~\cite{Du2006}.
% In particular typical AD regions, like the parietal lobe~\cite{}.
Similarly, when trained to detect FTD, \ourmod\ primarily considers the hallmark regions of this dementia type, the frontal and temporal lobe, see \Cref{fig:meanActivations}~(b). Additionally, we found activation in the inferior parietal lobe of the right hemisphere, which is further in line with findings about FTD~\cite{Du2006}. However, the temporal lobe received a comparably low weight in our FTD experiments. A reason could be that we do not explicitly enforce the model to consider \emph{all} discriminative regions; the model can learn to focus on the most relevant ones, therefore fostering easy interpretability.

\begin{figure}[t]
    \centering
    \includegraphics[width=\textwidth]{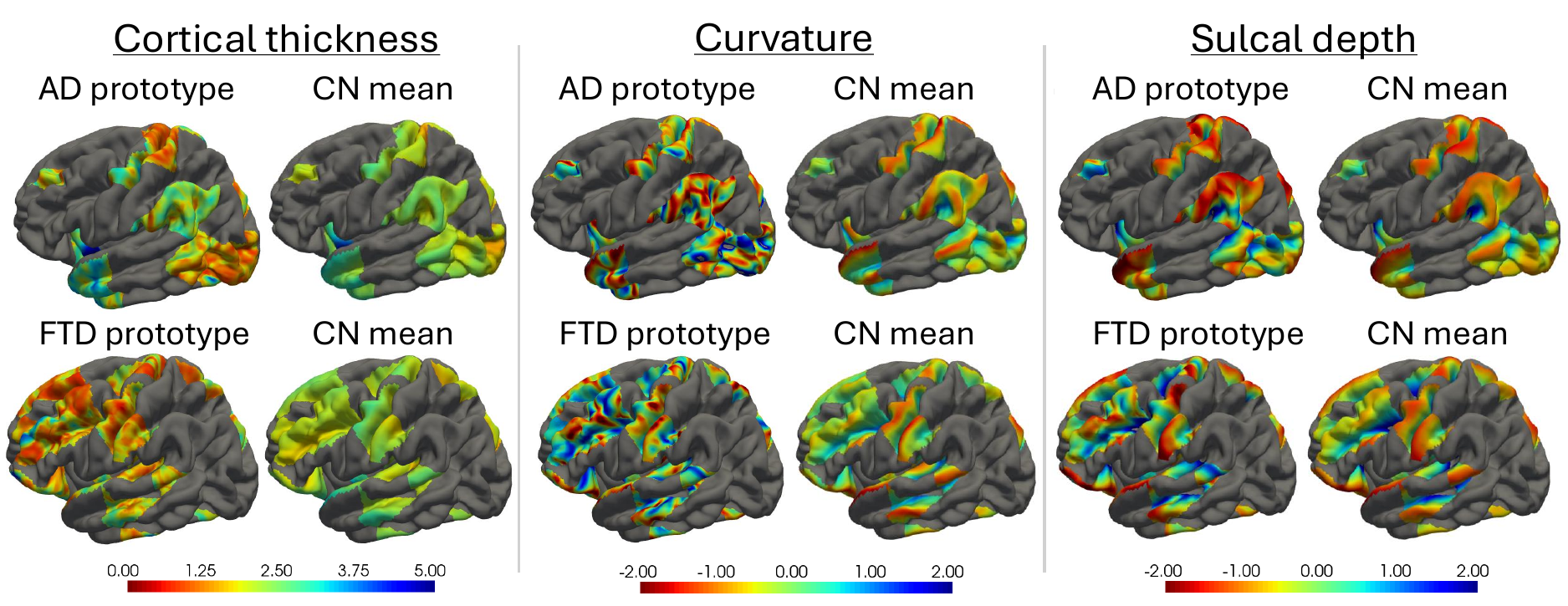}
    \caption{Prototypical disease-specific input features, i.e., prototypes, learned by \ourmod. We also show the mean features for the CN group in the training set for comparison. 
    Gray areas indicate regions that are ignored by the model~($w_i=0$, cf.~\Cref{sec:proto}). The visualization is based on the FsAverage template.
    }
    \label{fig:frankenstein}
\end{figure}

\noindent
\textbf{Individual level.}
On the individual level, \ourmod's explanations can aid in assessing the trustworthiness of a prediction. We depict patch-wise similarities between individual samples from the test set and the prototypes in \Cref{fig:meanActivations}~(c) and \Cref{fig:meanActivations}~(d). We observe that individual explanations for correctly predicted dementia cases predominantly follow the pattern observed at the group level. However, the individual explanation deviates considerably from the expected pattern in the case of wrong prediction. Consequently, users are enabled to question the model's predictions and conduct a manual review of the diagnosis when discrepancies arise. This rigorous assessment is essential for selecting optimal treatment strategies, especially in ambiguous or borderline cases.

\noindent
\textbf{Prototypes.}
By design, the predictions of \ourmod\ refer to the learned prototypes presented in \Cref{fig:frankenstein}. Prototypical patches are ``stitched together'' from different individuals; hence, they represent prototypical disease patterns that a single case could hardly cover. 
Compared to the average cortical thickness in CN cases, we observe a clear thinning in the learned prototypes. This is expected since atrophy in cortical gray matter is characteristic of the neurodegenerative process in both types of dementia~\cite{Du2006}. 
The association of curvature and sulcal depth with dementia is less extensively studied in the literature~\cite{Du2006,Singh2006corticalthinningalzheimers} and we did not find considerable differences apart from smoothing effects from taking the mean of the CN group. Yet, they provide a comprehensive picture of the prototypical anatomy used for the decision-making in our model. Importantly, we found the prototypes to be relatively robust to random weight initialization, achieving an overlap of 76.3\% and 71.7\% for the five AD and FTD models, respectively.

\section{Conclusion}
In conclusion, we introduced X-SiT, an interpretable model for surface-based detection of dementia based on high-resolution cortical features derived from structural MRI. Our results demonstrate that X-SiT matches the accuracy of state-of-the-art non-interpretable models in the field. Moreover, the explanations provided by X-SiT elucidate its decision-making process and help identify failure cases, thereby enhancing its usability and trustworthiness in clinical settings.

\begin{credits}
\subsubsection{\ackname} This research was partially supported by the German Research Foundation (DFG, No. 460880779). We gratefully acknowledge the computational resources provided by the Leibniz Supercomputing Centre (www.lrz.de).
This preprint has not undergone any post-submission improvements or corrections. The Version of Record of this contribution is published in Medical Image Computing and Computer Assisted Intervention --- MICCAI 2025, and is available online at \url{https://doi.org/10.1007/978-3-032-05185-1_65}.

\end{credits}

\bibliographystyle{splncs04}
\bibliography{references.bib}

\begin{thebibliography}{10}
\providecommand{\url}[1]{\texttt{#1}}
\providecommand{\urlprefix}{URL }
\providecommand{\doi}[1]{https://doi.org/#1}

\bibitem{Azcona2020interpretation}
Azcona, E.A., Besson, P., Wu, Y., Punjabi, A., Martersteck, A., Dravid, A., Parrish, T.B., Bandt, S.K., Katsaggelos, A.K.: Interpretation of brain morphology in association to alzheimer’s disease dementia classification using graph convolutional networks on triangulated meshes. In: Shape in Medical Imaging. p. 95–107 (2020)

\bibitem{deChastelaine2023cthCognitive}
de~Chastelaine, M., Srokova, S., Hou, M., Kidwai, A., Kafafi, S.S., Racenstein, M.L., Rugg, M.D.: Cortical thickness, gray matter volume, and cognitive performance: a crosssectional study of the moderating effects of age on their interrelationships. Cerebral Cortex  \textbf{33}(10),  6474–6485 (2023)

\bibitem{chen2019looks}
Chen, C., Li, O., Tao, D., Barnett, A., Rudin, C., Su, J.K.: This looks like that: deep learning for interpretable image recognition. Advances in neural information processing systems  \textbf{32} (2019)

\bibitem{Cheng2022spherical}
Cheng, J., Zhang, X., Zhao, F., Wu, Z., Yuan, X., Gilmore, J.H., Wang, L., Lin, W., Li, G.: Spherical transformer on cortical surfaces. In: Machine Learning in Medical Imaging. p. 406–415 (2022)

\bibitem{Chouliaras2023differentialDiagnosis}
Chouliaras, L., O’Brien, J.T.: The use of neuroimaging techniques in the early and differential diagnosis of dementia. Molecular Psychiatry  \textbf{28}(10),  4084–4097 (2023)

\bibitem{dahan22sit}
Dahan, S., Fawaz, A., Williams, L.Z.J., Yang, C., Coalson, T.S., Glasser, M.F., Edwards, A.D., Rueckert, D., Robinson, E.C.: Surface vision transformers: Attention-based modelling applied to cortical analysis. In: Medical Imaging with Deep Learning. Proceedings of Machine Learning Research, vol.~172, pp. 282--303 (2022)

\bibitem{Dale1999corticalI}
Dale, A.M., Fischl, B., Sereno, M.I.: Cortical surface-based analysis i. NeuroImage  \textbf{9}(2),  179–194 (1999)

\bibitem{dosovitskiy2021vit}
Dosovitskiy, A., Beyer, L., Kolesnikov, A., et~al.: An image is worth 16x16 words: Transformers for image recognition at scale. In: International Conference on Learning Representations (2021)

\bibitem{Du2006}
Du, A.T., Schuff, N., Kramer, J.H., Rosen, H.J., Gorno-Tempini, M.L., Rankin, K., Miller, B.L., Weiner, M.W.: Different regional patterns of cortical thinning in alzheimer’s disease and frontotemporal dementia. Brain  \textbf{130}(4),  1159–1166 (2006)

\bibitem{fischlFreeSurfer2012}
Fischl, B.: {{FreeSurfer}}. NeuroImage  \textbf{62}(2),  774--781 (2012)

\bibitem{Fischl1999corticalII}
Fischl, B., Sereno, M.I., Dale, A.M.: Cortical surface-based analysis ii. NeuroImage  \textbf{9}(2),  195–207 (1999)

\bibitem{gallee2023interpretable}
Gall{\'e}e, L., Beer, M., G{\"o}tz, M.: Interpretable medical image classification using prototype learning and privileged information. In: International Conference on Medical Image Computing and Computer-Assisted Intervention. pp. 435--445 (2023)

\bibitem{Gong2019spiral}
Gong, S., Chen, L., Bronstein, M., Zafeiriou, S.: Spiralnet++: A fast and highly efficient mesh convolution operator. In: 2019 IEEE/CVF International Conference on Computer Vision Workshop (ICCVW). p. 4141–4148 (2019)

\bibitem{Gopinath2022gcn}
Gopinath, K., Desrosiers, C., Lombaert, H.: Learnable pooling in graph convolutional networks for brain surface analysis. IEEE Transactions on Pattern Analysis and Machine Intelligence  \textbf{44}(2),  864–876 (2022)

\bibitem{jeyakumar2020can}
Jeyakumar, J.V., Noor, J., Cheng, Y.H., Garcia, L., Srivastava, M.: How can i explain this to you? an empirical study of deep neural network explanation methods. Advances in Neural Information Processing Systems  \textbf{33},  4211--4222 (2020)

\bibitem{kashefi2023explainability}
Kashefi, R., Barekatain, L., Sabokrou, M., Aghaeipoor, F.: Explainability of vision transformers: A comprehensive review and new perspectives. arxiv. arXiv preprint arXiv:2311.06786  (2023)

\bibitem{kim22vitnet}
Kim, S., Nam, J., Ko, B.C.: {V}i{T}-{N}e{T}: Interpretable vision transformers with neural tree decoder. In: International Conference on Machine Learning. Proceedings of Machine Learning Research, vol.~162, pp. 11162--11172 (2022)

\bibitem{kim2023help}
Kim, S.S., Watkins, E.A., Russakovsky, O., Fong, R., Monroy-Hern{\'a}ndez, A.: "help me help the ai": Understanding how explainability can support human-ai interaction. In: Conference on Human Factors in Computing Systems. pp. 1--17 (2023)

\bibitem{komorowski2023towards}
Komorowski, P., Baniecki, H., Biecek, P.: Towards evaluating explanations of vision transformers for medical imaging. In: Proceedings of the IEEE/CVF conference on computer vision and pattern recognition. pp. 3726--3732 (2023)

\bibitem{loshchilov2018decoupled}
Loshchilov, I., Hutter, F.: Decoupled weight decay regularization. In: International Conference on Learning Representations (2019)

\bibitem{mohammadjafari2021using}
Mohammadjafari, S., Cevik, M., Thanabalasingam, M., Basar, A.: Using protopnet for interpretable alzheimer's disease classification. In: Canadian AI (2021)

\bibitem{PahuddeMortanges2024orchestrating}
Pahud~de Mortanges, A., Luo, H., Shu, S.Z., Kamath, A., Suter, Y., Shelan, M., P\"{o}llinger, A., Reyes, M.: Orchestrating explainable artificial intelligence for multimodal and longitudinal data in medical imaging. npj Digital Medicine  \textbf{7}(1) (2024)

\bibitem{Munroe2024interpretableNeuroReview}
Munroe, L., da~Silva, M., Heidari, F., Grigorescu, I., Dahan, S., Robinson, E.C., Deprez, M., So, P.W.: Applications of interpretable deep learning in neuroimaging: A comprehensive review. Imaging Neuroscience  \textbf{2},  1–37 (2024)

\bibitem{nauta2023pipnet}
Nauta, M., Schlötterer, J., van Keulen, M., Seifert, C.: Pip-net: Patch-based intuitive prototypes for interpretable image classification. Proceedings of the IEEE/CVF Conference on Computer Vision and Pattern Recognition  (2023)

\bibitem{nguyen2021effectiveness}
Nguyen, G., Kim, D., Nguyen, A.: The effectiveness of feature attribution methods and its correlation with automatic evaluation scores. Advances in Neural Information Processing Systems  \textbf{34},  26422--26436 (2021)

\bibitem{Nyul2000standardization}
Nyul, L., Udupa, J., Zhang, X.: New variants of a method of mri scale standardization. IEEE Transactions on Medical Imaging  \textbf{19}(2),  143–150 (2000)

\bibitem{Pang2023geometricConstraints}
Pang, J.C., Aquino, K.M., Oldehinkel, M., Robinson, P.A., Fulcher, B.D., Breakspear, M., Fornito, A.: Geometric constraints on human brain function. Nature  \textbf{618}(7965),  566–574 (2023)

\bibitem{rudin2019stop}
Rudin, C.: Stop explaining black box machine learning models for high stakes decisions and use interpretable models instead. Nature machine intelligence  \textbf{1}(5),  206--215 (2019)

\bibitem{rudin2022interpretable}
Rudin, C., Chen, C., Chen, Z., Huang, H., Semenova, L., Zhong, C.: Interpretable machine learning: Fundamental principles and 10 grand challenges. Statistic Surveys  \textbf{16},  1--85 (2022)

\bibitem{Singh2006corticalthinningalzheimers}
Singh, V., Chertkow, H., Lerch, J.P., Evans, A.C., Dorr, A.E., Kabani, N.J.: Spatial patterns of cortical thinning in mild cognitive impairment and alzheimer’s disease. Brain  \textbf{129}(11),  2885–2893 (2006)

\bibitem{Spitzer2022interpretable}
Spitzer, H., Ripart, M., Whitaker, K., et~al.: Interpretable surface-based detection of focal cortical dysplasias: a multi-centre epilepsy lesion detection study. Brain  \textbf{145}(11),  3859–3871 (2022)

\bibitem{tibshirani1996regression}
Tibshirani, R.: Regression shrinkage and selection via the lasso. Journal of the Royal Statistical Society Series B: Statistical Methodology  \textbf{58}(1),  267--288 (1996)

\bibitem{wolf2023don}
Wolf, T.N., P{\"o}lsterl, S., Wachinger, C.: Don’t panic: Prototypical additive neural network for interpretable classification of alzheimer’s disease. In: International Conference on Information Processing in Medical Imaging. pp. 82--94 (2023)

\bibitem{Yang2022unbox}
Yang, G., Ye, Q., Xia, J.: Unbox the black-box for the medical explainable ai via multi-modal and multi-centre data fusion: A mini-review, two showcases and beyond. Information Fusion  \textbf{77},  29–52 (2022)

\bibitem{Zhao2019sunet}
Zhao, F., Xia, S., Wu, Z., Duan, D., Wang, L., Lin, W., Gilmore, J.H., Shen, D., Li, G.: Spherical U-Net on Cortical Surfaces: Methods and Applications, p. 855–866. Springer International Publishing (2019)

\end{thebibliography}

\end{document}